\documentclass[reprint,amsmath,amssymb]{revtex4-1}

\usepackage{graphicx}
\usepackage{color}

\begin{document}


\title{Bias voltage effects on tunneling magnetoresistance in Fe/MgAl${}_2$O${}_4$/Fe(001) junctions: Comparative study with Fe/MgO/Fe(001) junctions}


\author{Keisuke Masuda}
\affiliation{Research Center for Magnetic and Spintronic Materials, National Institute for Materials Science (NIMS), 1-2-1 Sengen, Tsukuba 305-0047, Japan}

\author{Yoshio Miura}
\affiliation{Research Center for Magnetic and Spintronic Materials, National Institute for Materials Science (NIMS), 1-2-1 Sengen, Tsukuba 305-0047, Japan}
\affiliation{Kyoto Institute of Technology, Electrical Engineering and Electronics, Kyoto 606-8585, Japan}
\affiliation{Center for Materials Research by Information Integration, National Institute for Materials Science (NIMS), 1-2-1 Sengen, Tsukuba 305-0047, Japan}
\affiliation{Center for Spintronics Research Network (CSRN), Graduate School of Engineering Science, Osaka University, Machikaneyama 1-3, Toyonaka, Osaka 560-8531, Japan}


\date{\today}

\begin{abstract}
We investigate bias voltage effects on the spin-dependent transport properties of Fe/MgAl${}_2$O${}_4$/Fe(001) magnetic tunneling junctions (MTJs) by comparing them with those of Fe/MgO/Fe(001) MTJs. By means of the nonequilibrium Green's function method and the density functional theory, we calculate bias voltage dependences of magnetoresistance (MR) ratios in both the MTJs. We find that in both the MTJs, the MR ratio decreases as the bias voltage increases and finally vanishes at a critical bias voltage $V_{\rm c}$. We also find that the critical bias voltage $V_{\rm c}$ of the MgAl${}_2$O${}_4$-based MTJ is clearly larger than that of the MgO-based MTJ. Since the in-plane lattice constant of the Fe/MgAl${}_2$O${}_4$/Fe(001) supercell is twice that of the Fe/MgO/Fe(001) one, the Fe electrodes in the MgAl${}_2$O${}_4$-based MTJs have an identical band structure to that obtained by folding the Fe band structure of the MgO-based MTJs in the Brillouin zone of the in-plane wave vector. We show that such a difference in the Fe band structure is the origin of the difference in the critical bias voltage $V_{\rm c}$ between the MgAl${}_2$O${}_4$- and MgO-based MTJs.
\end{abstract}


\maketitle

\section{introduction}
Magnetic tunneling junctions (MTJs), in which ferromagnetic electrodes are separated by an insulating barrier, are key systems for realizing high-performance spintronic devices such as nonvolatile magnetic random access memories (MRAMs) and read heads of hard disk drives (HDDs). Various combinations of ferromagnets and insulating barriers have been tested in order to achieve high magnetoresistance (MR) ratios in MTJs. Although amorphous alumina barriers gave only low MR ratios \cite{1995Miyazaki-JMMM,1995Moodera-PRL}, the use of crystalline MgO barriers increased MR ratios to $300\%$ at low temperature and $200\%$ at room temperature \cite{2004Parkin-NatMat,2004Yuasa-NatMat}. Such effectiveness of MgO barriers was originally proposed in the pioneering theoretical works by Butler {\it et al.} \cite{2001Butler-PRB} and Mathon {\it et al.} \cite{2001Mathon-PRB}. In MgO barriers, the wave functions in the $\Delta_1$ states ($s$, $p_z$, and $d_{3 z^2 - r^2}$ states) have the slowest decay as evanescent waves \cite{2001Butler-PRB}. Furthermore, since typical ferromagnetic electrodes (Fe, Co, and some Heusler compounds) have half-metallicity in the $\Delta_1$ states, the majority-spin electrons in the $\Delta_1$ states are selectively transmitted from one electrode to another through the MgO barrier, which is the origin of the high MR ratios observed in MgO-based MTJs \cite{2001Butler-PRB,2001Mathon-PRB}. Namely, the MgO barrier plays the role of a spin filter in MTJs.

The spinel oxides, $XY_2$O$_4$, including more than 200 compounds, can also be insulating barriers of MTJs. In particular, MgAl$_2$O$_4$ has been studied from both theoretical and experimental points of view. Like in MgO, the wave functions in the $\Delta_1$ states decay most slowly in MgAl$_2$O$_4$ \cite{2012Zhang-APL,2012Miura-PRB}. On the other hand, the first-principles-based transport calculations by Miura {\it et al.} predicted that Fe/MgAl$_2$O$_4$/Fe(001) MTJs have much smaller MR ratios (160$\%$) than the Fe/MgO/Fe(001) MTJs (1600$\%$) owing to the ``band-folding'' effect \cite{2012Miura-PRB}. Since the in-plane lattice constant of the Fe/MgAl$_2$O$_4$/Fe(001) supercell is twice as long as that of bcc Fe, the original bands of bcc Fe are folded and not only majority-spin bands but also minority-spin bands cross the Fermi level in the $\Delta_1$ states \cite{2012Miura-PRB}. This implies the absence of the half-metallicity in the $\Delta_1$ states, leading to small MR ratios. Such a theoretical prediction is consistent with early-stage experiments on Fe/MgAl$_2$O$_4$/Fe(001) MTJs \cite{2010Sukegawa-APL}, where an MR ratio of 165 $\%$ was observed at low temperature for MgAl$_2$O$_4$ made from Mg/Al bilayers. However, in subsequent experiments, Sukegawa {\it et al.} clarified that MgAl$_2$O$_4$ made from Mg-Al alloys gives much higher MR ratios over 300$\%$ at low temperature \cite{2012Sukegawa-PRB}, which exceeds the above theoretical limit. They also showed that the MgAl$_2$O$_4$ made from Mg-Al alloys has disordered cation sites (referred to as the cation-disorder MgAl$_2$O$_4$) by transmission electron microscopy (TEM), giving a lattice constant close to that of bcc Fe [see Figs. 1(a) and 1(b) of Ref. \cite{2012Sukegawa-PRB} for details]. Therefore, the folding of the Fe bands does not occur in the cation disordered MgAl$_{2}$O$_{4}$ and hence the MR ratios can be large. Most recently, an even higher MR ratio of 436$\%$ at $3\,{\rm K}$ (245$\%$ at $297\,{\rm K}$) was achieved for an MTJ with the cation-disorder MgAl$_2$O$_4$ prepared by direct sputtering from a sintered target \cite{2016Belmoubarik-APL}.

One of the significant features of the MgAl$_2$O$_4$-based MTJs is the good lattice matching between MgAl$_2$O$_4$ and ferromagnetic electrodes. The lattice mismatches of MgAl$_2$O$_4$ with typical ferromagnets (e.g., Fe, Co, and the Heusler alloy Co$_2$FeAl) are around 0.1$\%$, which are approximately one order smaller than those of the MgO-based MTJs. Furthermore, the lattice constant of MgAl$_2$O$_4$ can be tuned by changing the Mg/Al composition rate, enabling good lattice matching with other ferromagnets with large perpendicular magnetic anisotropy, e.g., $D0_{22}$ Mn$_{3}$Ga \cite{2009Wu-APL,2011Mizukami-APL}, $D0_{22}$ Mn$_{3}$Ge \cite{2012Kurt-APL,2013Mizukami-APEX}, $L1_{0}$ MnGa \cite{2011Mizukami-APL}, and $L1_{0}$ FePt \cite{1995Klemmer-SMM}, which are advantageous for MRAM applications from the viewpoint of thermal stability. Another important feature of the MgAl$_2$O$_4$-based MTJs is the robustness of the MR ratios to the applied bias voltage. To estimate the robustness quantitatively, we often use $V_{\rm half}$, defined as the bias voltage where an MR ratio becomes half of the zero-bias value. In the MgAl$_2$O$_4$-based MTJs, quite high values of $V_{\rm half}$, over 1\,V, which is about twice that for the MgO-based MTJs \cite{2004Yuasa-NatMat,2005Djayaprawira-APL,2009Wang-APL}, have been observed \cite{2010Sukegawa-APL,2016Belmoubarik-APL,2013Sukegawa-APL,2014Sukegawa-APL}. Such a high $V_{\rm half}$, leading to a high value of the output voltage, is preferable for device applications. Therefore, the MgAl$_2$O$_4$-based MTJ is one of the most suitable systems for investigating bias voltage effects on spin-dependent transport properties. However, to the best of our knowledge, no theoretical study has been done on this issue.

In this work, we theoretically study bias voltage effects on the electronic structures and spin-dependent transport properties of Fe/MgAl$_2$O$_4$/Fe(001) MTJs. In particular, we focus on the bias voltage dependence of the MR ratio, which is compared with that of the Fe/MgO/Fe(001) MTJs. First, we investigate zero-bias transport properties of both the MgAl$_2$O$_4$- and MgO-based MTJs, which help in understanding the bias voltage effects. We clarify which conductive channel is dominant at each energy level for both the MTJs. By using the nonequilibrium Green's function method in combination with the density functional theory, we calculate the bias voltage dependencies of the currents, MR ratios, and transmittances in both the MTJs. We found that in both the MTJs, the MR ratio decreases with increasing the bias voltage $V$ and eventually vanishes at a critical value $V_{\rm c}$. We also found that the MgAl$_2$O$_4$-based MTJ has a larger $V_{\rm c}$ than the MgO-based one. As mentioned above, in Fe/MgAl$_2$O$_4$/Fe(001) MTJs, the original bands of Fe electrodes are folded owing to the large supercell. We clarify that this band folding effect is the origin of the larger $V_{\rm c}$ in the MgAl$_2$O$_4$-based MTJs.

\section{\label{structure optimization} structure optimization}
\begin{figure}
\includegraphics[width=8cm]{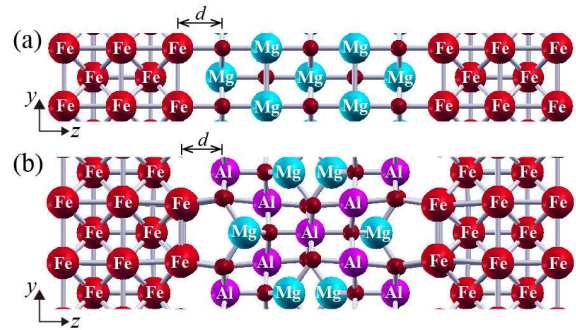}
\caption{\label{supercells} Supercells of (a) Fe(5)/MgO(5)/Fe(5) and (b) Fe(5)/MgAl$_2$O$_4$(9)/Fe(5), optimized by the procedures of Sec. \ref{structure optimization}.}
\end{figure}
To properly estimate transport properties of MTJs, we need to optimize atomic configurations in the supercells used in our transport calculations. We prepared the supercells Fe(5)/MgO(5)/Fe(5) and Fe(5)/MgAl$_2$O$_4$(9)/Fe(5) shown in Fig. \ref{supercells}, where each number represents the layer number of each compound. In a previous study \cite{2012Miura-PRB}, one of us confirmed that a layer consisting of Al and O atoms is energetically favored as the termination layer of MgAl$_2$O$_4$ in Fe/MgAl$_2$O$_4$(001). Moreover, it was also confirmed that the interfacial atomic configuration in which O atoms are on top of Fe atoms has the lowest energy. Considering these facts, we adopted the atomic configuration shown in Fig. \ref{supercells}(b) for the Fe/MgAl$_2$O$_4$/Fe(001) supercell. The in-plane lattice constants of the MgO- and MgAl$_2$O$_4$-based supercells were fixed to one and two times that of bcc Fe (2.866\,{\AA}), respectively. As initial atomic configurations, we utilized the bulk atomic positions of Fe, MgO, and MgAl$_2$O$_4$. Note that the distance $d$ between the electrode and the barrier (see Fig.~\ref{supercells}) also needs to be optimized in addition to the atomic configurations. First, for some fixed values of $d$, we calculated the ground-state energy $E_0$ of the supercell relaxing the atomic positions. We next obtained a $d$ versus $E_0$ curve with the help of the spline interpolation. From the curve, the distance $d_1$ with a minimum value of $E_0$ was determined. Finally, we calculated the ground-state energy of the supercell at $d_1$, from which we obtained the optimized atomic positions in the supercell. All these calculations were done using the density functional theory in combination with the generalized gradient approximation, implemented in the Vienna $ab$ $initio$ simulation program (VASP) \cite{1996Kresse-PRB,1999Kresse-PRB}. We used $20\times20\times3$ and $10\times10\times3$ ${\bf k}$-point meshes in the MgO and MgAl$_2$O$_4$ cases, respectively. We also assumed that the spins of all Fe atoms in the supercell align parallel to each other in both cases. As a result of the calculations, the distance between the electrode and the barrier was determined to be 2.2\,{\AA} and 2.0\,{\AA} in the MgO and MgAl$_2$O$_4$ cases, respectively.

\section{\label{zero-bias case} zero-bias case}
\begin{figure*}
\includegraphics[width=18cm]{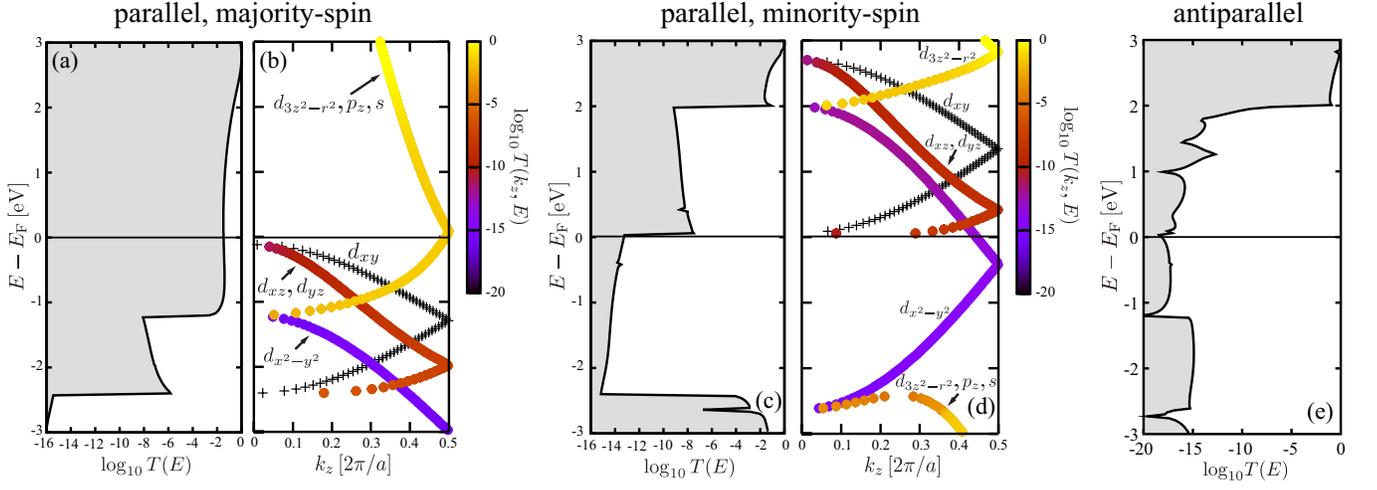}
\caption{\label{mgo_cond-band} Majority-spin transmittance at ${\bf k}_{\|} = (0,0)$ in Fe/MgO/Fe(001) MTJs with parallel magnetization of electrodes. (a) The energy dependence of the total transmittance in the [001] direction. (b) The band-resolved transmittance in the [001] direction. At each point of the Fe majority-spin bands in the [001] direction, the logarithmic magnitude of the resolved transmittance is shown as a color in the color palette. In each band, constituent orbitals are listed from largest to smallest contribution. (c) and (d) The same as (a) and (b) for minority-spin channel, respectively. (e) The same as (a) for antiparallel magnetization of electrodes.}
\end{figure*}
For each of the MgO- and MgAl$_2$O$_4$-based MTJs with zero-bias voltage, we considered the quantum open system composed of the supercell (obtained in Sec. \ref{structure optimization}) attached to the left and right semi-infinite electrodes of Fe atoms.  We calculated the transmittance of each quantum open system using the quantum code ESPRESSO \cite{Baroni}. This code can analyze the band-resolved transmittance at zero-bias voltage, which gives valuable information to understand bias voltage effects on the transport properties of MTJs. First, we obtained the self-consistent potential of the quantum open system by means of the density functional theory and the generalized gradient approximation, where a $10\times10\times1$ ${\bf k}$-point mesh and Methfessel--Paxton smearing with the broadening parameter 0.01 Ry were used. We also set the cutoff energies for the wave functions and the charge density to 30 and 300 Ry, respectively. Owing to the two-dimensional periodicity of our systems, the scattering states can be classified by an in-plane wave vector ${\bf k}_{\|} = (k_{x},k_{y})$. For each ${\bf k}_{\|}$ and spin index, we solved the scattering equations derived under the condition that the wave function and its derivative of the supercell are connected to those of the Fe electrodes \cite{1999Choi-PRB,2004Smogunov-PRB}. In this section, we focus only on the transmittance at ${\bf k}_{\|} = (0,0)$ since this component provides the dominant contribution to the total transmittance in coherent tunneling processes.
\begin{figure*}
\includegraphics[width=18cm]{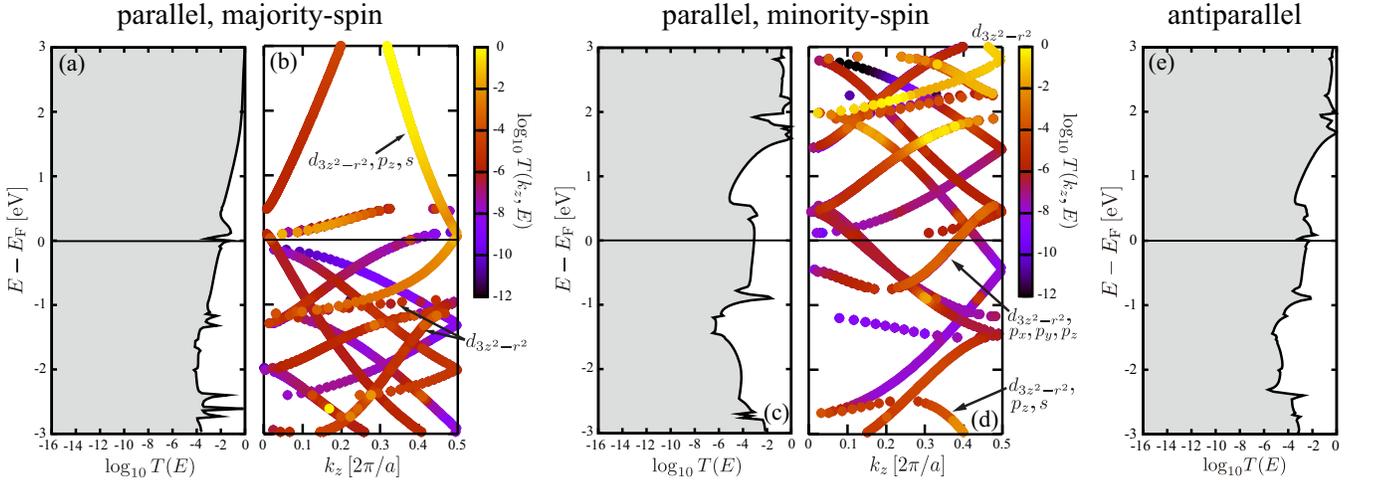}
\caption{\label{mao_cond-band} The same as Fig. \ref{mgo_cond-band} for Fe/MgAl$_2$O$_4$/Fe(001) MTJs.}
\end{figure*}

Figure~\ref{mgo_cond-band}(a) shows the energy dependence of the majority-spin transmittance in the MgO-based MTJ with parallel magnetization of Fe electrodes. We can see a clear three-step structure: $T(E) \gtrsim 10^{-2}$ for $E-E_{\rm F} \gtrsim -1.2\,{\rm eV}$, $10^{-6} \gtrsim T(E) \gtrsim 10^{-8}$ for $-1.2\,{\rm eV} \gtrsim E-E_{\rm F} \gtrsim -2.4\,{\rm eV}$, and $10^{-15} \gtrsim T(E)$ for $-2.4\,{\rm eV} \gtrsim E-E_{\rm F}$. To understand this behavior, we show the band-resolved transmittance in Fig.~\ref{mgo_cond-band}(b). By comparing Figs.~\ref{mgo_cond-band}(a) and \ref{mgo_cond-band}(b), we find that the drops in the transmittance in Fig.~\ref{mgo_cond-band}(a) occur at the band edges of the majority-spin Fe bands. In the highest energy region ($E-E_{\rm F} \gtrsim -1.2\,{\rm eV}$), the majority-spin Fe band mainly from the $d_{3z^2-r^2}$ state gives quite large transmittances. In the middle-energy region ($-1.2\,{\rm eV} \gtrsim E-E_{\rm F} \gtrsim -2.4\,{\rm eV}$), the band from the $d_{xz}$ and $d_{yz}$ states provides the dominant contribution to the total transmittance although the presence of another band from the $d_{x^2-y^2}$ state. The lowest energy region ($-2.4\,{\rm eV} \gtrsim E-E_{\rm F}$) has only the band from the $d_{x^2-y^2}$ state, which gives quite small transmittances. A previous theoretical study \cite{2001Butler-PRB} showed that $\Delta_1$ ($d_{3z^2-r^2}$, $p_z$, $s$), $\Delta_5$ ($d_{xz}$, $d_{yz}$, $p_x$, $p_y$), and $\Delta_2$ ($d_{x^2-y^2}$, $d_{xy}$) states of Fe have the first, second, and third slowest decays as evanescent waves in the MgO barrier, respectively, which is consistent with our present results. Figure~\ref{mgo_cond-band}(c) shows the energy dependence of the minority-spin transmittance. As seen from the band-resolved transmittance in Fig.~\ref{mgo_cond-band}(d), the $d_{3z^2-r^2}$ state gives quite large transmittances in the high- and low-energy regions away from the Fermi level. The second largest transmittances are given by the $d_{xz}$ and $d_{yz}$ states in the region of $2\,{\rm eV} \gtrsim E-E_{\rm F} \gtrsim 0\,{\rm eV}$. Below the Fermi level ($0\,{\rm eV} \gtrsim E-E_{\rm F} \gtrsim -2.4\,{\rm eV}$), the $d_{x^2-y^2}$ state provides the smallest transmittances. These are also consistent with the previous theoretical results \cite{2001Butler-PRB}. As is clear from Figs.~\ref{mgo_cond-band}(b) and \ref{mgo_cond-band}(d), the Fe electrodes have half-metallicity in the $\Delta_1$ states ($d_{3z^2-r^2}$-based states), which gives a large spin-dependent transmittance at the Fermi level, namely, a relatively high MR ratio \cite{2001Butler-PRB,2001Mathon-PRB}. In Fig. \ref{mgo_cond-band}(e), we show the transmittance in the antiparallel magnetization state of Fe electrodes, where the left and right electrodes have opposite directions of magnetization. In this case, tunneling of electrons occurs between majority- and minority-spin bands. We see that the transmittance has large values $T(E) \gtrsim 10^{-2}$ for $E-E_{\rm F} \gtrsim 2\,{\rm eV}$, in which both the majority- and minority-spin states have $d_{3z^2-r^2}$-based band giving large transmittance as shown in Figs. \ref{mgo_cond-band}(b) and \ref{mgo_cond-band}(d). We also see that other energy regions have quite small transmittances $T(E) \lesssim 10^{-12}$ owing to the absence of the $d_{3z^2-r^2}$-based band in either the majority- or minority-spin state.

Next, we show in Fig.~\ref{mao_cond-band}(a) the energy dependence of the majority-spin transmittance in the MgAl$_2$O$_4$-based MTJ with parallel magnetization of Fe electrodes. We also show the band-resolved transmittance in Fig.~\ref{mao_cond-band}(b), where a larger number of majority-spin Fe bands are seen than for the MgO case [Fig.~\ref{mgo_cond-band}(b)]. This is due to the band-folding effect pointed out in previous studies \cite{2012Miura-PRB,2012Sukegawa-PRB}. Since the in-plane lattice constant of the Fe/MgAl$_2$O$_4$/Fe(001) supercell is twice as long as that of bcc Fe [see Fig.~\ref{supercells}(b)], the original Fe bands are folded in the ${\bf k}_{\|}$ space, resulting in the large number of bands. In Fig.~\ref{mao_cond-band}(b), we find that the band mainly from the $d_{3z^2-r^2}$ state provides the dominant contribution to the total transmittance in the energy region above the Fermi level. On the other hand, below the Fermi level, the total transmittance has contributions from various bands, but the $d_{3z^2-r^2}$-based bands play the dominant role \cite{note}. Previous theoretical studies on Fe/MgAl$_2$O$_4$/Fe(001) MTJs \cite{2012Zhang-APL,2012Miura-PRB} have shown that the $d_{3z^2-r^2}$-based Fe state has the slowest decay in the MgAl$_2$O$_4$ barrier, which is consistent with our results. Note here again that in Fig.~\ref{mao_cond-band}(b), the $d_{3z^2-r^2}$-based bands exist in the whole energy region around the Fermi level because of the band-folding effect. This is clearly different from the MgO case [Fig.~\ref{mgo_cond-band}(b)] and is the origin of the smooth energy dependence of the majority-spin transmittance shown in Fig.~\ref{mao_cond-band}(a). Figure~\ref{mao_cond-band}(c) shows the energy dependence of the minority-spin transmittance. From Fig.~\ref{mao_cond-band}(d), we see that the $d_{3z^2-r^2}$-based bands provide the dominant contribution to the total transmittance. We also find that the half-metallicity in the $d_{3z^2-r^2}$-based bands is absent, unlike the MgO-based MTJs, because the folded minority-spin band mainly from the $d_{3z^2-r^2}$ state crosses the Fermi level. Here, it is worthy to note that the minority-spin $d_{3z^2-r^2}$-based folded band crossing the Fermi level of bcc Fe additionally includes the $p_x$ and $p_y$ states without the total symmetry about the $z$-axis rotation [see Fig.~\ref{mao_cond-band}(d)]. This is in contrast to the majority-spin case [Fig.~\ref{mao_cond-band}(b)], where the $d_{3z^2-r^2}$-based band crossing the Fermi level includes only the $d_{3z^2-r^2}$, $p_z$, and $s$ states. Since the mixing of the $p_x$ and $p_y$ states affects the spatial distribution of the wave function of the folded minority-spin band, the majority- and minority-spin states in Fe electrodes are expected to have different overlaps with the wave functions of the $\Delta_1$ evanescent state of the MgAl$_2$O$_4$ barrier. In the case of parallel magnetization, the scattering wave function of the $\Delta_1$ evanescent minority-spin state shows a strong reduction at both sides of the Fe/MgAl$_2$O$_4$ interface, due to the partial overlaps between the folded minority-spin state of Fe (including $d_{3z^2-r^2}$, $p_x$, and $p_y$) and the evanescent $\Delta_1$ state of MgAl$_2$O$_4$ [see Fig. 6(b) of Ref.~\cite{2012Miura-PRB}]. This leads to the small transmittance compared with the majority-spin state, in spite of the same decay rate of the $\Delta_1$ evanescent state for the majority- and minority-spin channels in the barrier region. On the other hand, in the case of the antiparallel magnetization, significant reduction of the scattering wave function at the Fe/MgAl$_2$O$_4$ interface occurs only at one side of the junction, because of the complete matching between the majority-spin $\Delta_1$ state of Fe and the evanescent $\Delta_1$ state of MgAl$_2$O$_4$ at the other side of the junction. Thus, the transmittance in the antiparallel magnetization state of Fe/MgAl$_2$O$_4$/Fe(001) MTJs is larger than that in the minority-spin state in the case of the parallel magnetization [see Figs. \ref{mao_cond-band}(c) and \ref{mao_cond-band}(e)], which is clearly different from the case of Fe/MgO/Fe(001) MTJs.

\section{finite-bias case}
Transport calculations under finite-bias voltages were carried out using the nonequilibrium Green's function method in combination with the density functional theory, implemented in the Atomistix ToolKit package (ATK) \cite{ATK,2002Brandbyge-PRB,2002Soler-JPCM}. First, we constructed a Hamiltonian of the supercell attached to the left and right semi-infinite Fe electrodes, between which a finite voltage $V$ was applied. As the supercells, we used Fe(7)/MgO(5)/Fe(7) and Fe(7)/MgAl$_2$O$_4$(9)/Fe(7), optimized by the procedure of Sec.~\ref{structure optimization} \cite{remark}. In addition to the pseudopotential and exchange--correlation potential, the Hamiltonian also includes the Hartree potential determined from the electron density by solving the Poisson equation \cite{2002Brandbyge-PRB}.

From the Hamiltonian, we can obtain the following retarded Green's function matrix $G_{\sigma}(\epsilon)$ with spin $\sigma$ and energy $\epsilon$:
\begin{eqnarray}
G_{\sigma}(\epsilon)= \frac{1}{(\epsilon+i\delta)S_{\sigma}-H_{\sigma}-\Sigma_{L,\sigma}(\epsilon)-\Sigma_{R,\sigma}(\epsilon)},
\end{eqnarray}
where $S_{\sigma}$ and $H_{\sigma}$ are the overlap and Hamiltonian matrices, respectively. Here, $\Sigma_{L,\sigma}(\epsilon)$ and $\Sigma_{R,\sigma}(\epsilon)$ represent the self-energy matrices from the left and right electrodes, respectively. Using the Green's function matrix $G_{\sigma}(\epsilon)$, the density matrix $D_{\sigma} \equiv D^{L}_{\sigma}+D^{R}_{\sigma}$ can be calculated, where $D^{L}_{\sigma}$ and $D^{R}_{\sigma}$ are given by
\begin{eqnarray}
D^{L}_{\sigma}&=&\frac{1}{2\pi} \int G_{\sigma}(\epsilon) f(\epsilon-\mu_{L}) \Gamma^{L}_{\sigma}(\epsilon) G^{\dagger}_{\sigma}(\epsilon) \, d\epsilon, \\
D^{R}_{\sigma}&=&\frac{1}{2\pi} \int G_{\sigma}(\epsilon) f(\epsilon-\mu_{R}) \Gamma^{R}_{\sigma}(\epsilon) G^{\dagger}_{\sigma}(\epsilon) \, d\epsilon.
\end{eqnarray}
Here, $\mu_L$ ($\mu_R$) is the chemical potential in the left (right) electrode and $f(\epsilon)=1/(e^{\beta \epsilon}+1)$ with $\beta=1/k_{\rm B}T$ is the Fermi distribution function. In this work, we assume that electrons flow from the left to right electrode by a positive bias $V$, namely, $\mu_L-\mu_R=eV$. The matrices $\Gamma^{L}_{\sigma}(\epsilon)$ and $\Gamma^{R}_{\sigma}(\epsilon)$ are defined as
\begin{eqnarray}
\Gamma^{L}_{\sigma}(\epsilon)&=&i \left( \Sigma_{L,\sigma}(\epsilon)-\Sigma^{\dagger}_{L,\sigma}(\epsilon) \right), \\
\Gamma^{R}_{\sigma}(\epsilon)&=&i \left( \Sigma_{R,\sigma}(\epsilon)-\Sigma^{\dagger}_{R,\sigma}(\epsilon) \right).
\end{eqnarray}
From the density matrix $D_{\sigma}$, we obtain the electron density $n_{\sigma}({\bf r})$ for spin $\sigma$ through the following expression:
\begin{eqnarray}
n_{\sigma}({\bf r})=\sum_{\mu,\nu} \phi_{\mu}({\bf r})\, {\rm Re} \left[ \left(D_{\sigma}\right)_{\mu,\nu} \right] \phi_{\nu}({\bf r}),
\end{eqnarray}
where $\phi_{\mu}({\bf r})$ is a real basis function. All these procedures were repeated iteratively until the electron density converged. In these self-consistent calculations, we set the electron temperature (thermal smearing of the Fermi distribution) to 1200 K and the density mesh cutoff to 75 Hartree. For the calculations of Fe electrodes, we used a $7\times7\times50$ ${\bf k}$-point mesh. Such a large number of $k_{z}$ points is required to minimize the mismatch of the Fermi energy between the Fe electrode and scattering region including the barrier, where an open boundary condition is adopted to treat finite-bias voltages. We also utilized the double-$\zeta$ and single-$\zeta$ polarized basis sets for the MgO- and MgAl$_2$O$_4$-based MTJs, respectively. By using the Green's function matrix $G_{\sigma}(\epsilon)$ with the converged electron density $n_{\sigma}({\bf r})$, the current $I_{\sigma}(V)$ at zero temperature can be calculated as follows:
\begin{eqnarray}
I_{\sigma}(V)= \frac{e}{h} \int^{\mu_{L}}_{\mu_{R}}\, T_{\sigma}(\epsilon)\, d\epsilon,\label{eq:current}
\end{eqnarray}
where $T_{\sigma}(\epsilon)$ is the transmittance given by
\begin{eqnarray}
T_{\sigma}(\epsilon)={\rm Tr} \left[ \Gamma^{R}_{\sigma}(\epsilon) G^{\dagger}_{\sigma}(\epsilon)\, \Gamma^{L}_{\sigma}(\epsilon) G_{\sigma}(\epsilon) \right]\label{eq:transmittance}.
\end{eqnarray}
In such transport calculations, ${\bf k}_{\|}$-point samplings were performed using $101\times101$ and $151\times151$ points in the ${\bf k}_{\|}$ Brillouin zones for the MgO and MgAl$_2$O$_4$ cases, respectively.

\begin{figure}
\includegraphics[width=6cm]{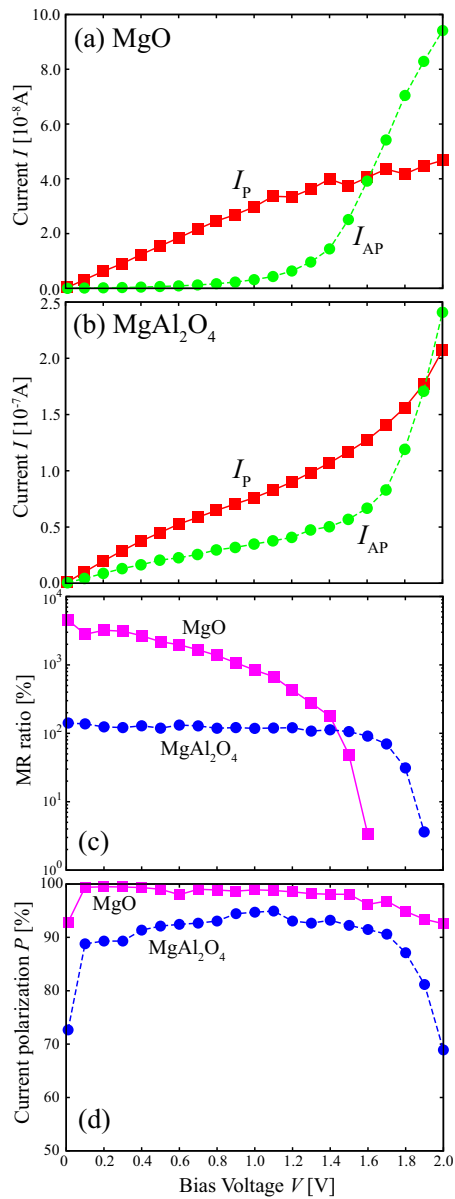}
\caption{\label{I-V_V-MR_V-P} The bias voltage dependences of currents in (a) Fe/MgO/Fe(001) and (b) Fe/MgAl$_2$O$_4$/Fe(001) MTJs. The bias voltage dependences of (c) MR ratios and (d) current polarizations in both the MTJs (see text for details).}
\end{figure}
\begin{figure}
\includegraphics[width=8.8cm]{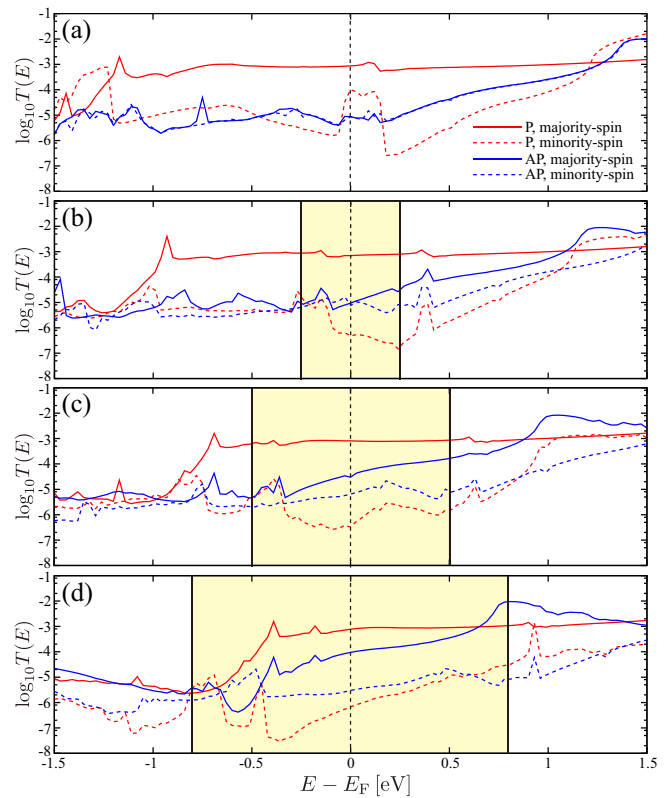}
\caption{\label{trans_MgO} Energy dependences of transmittance integrated over ${\bf k}_{\parallel}$ in Fe/MgO/Fe(001) MTJs for (a) $V=0.0\,{\rm V}$, (b) $V=0.5\,{\rm V}$, (c) $V=1.0\,{\rm V}$, and (d) $V=1.6\,{\rm V}$. In each panel, the red solid (dashed) curve corresponds to the majority- (minority)-spin transmittance in the case of parallel (P) magnetization of electrodes. The blue solid (dashed) curve represents the majority- (minority)-spin transmittance in the case of antiparallel (AP) magnetization of electrodes. The shaded area represents the integral region for the calculation of the current following Eq. (\ref{eq:current}).}
\end{figure}
\begin{figure}
\includegraphics[width=8.8cm]{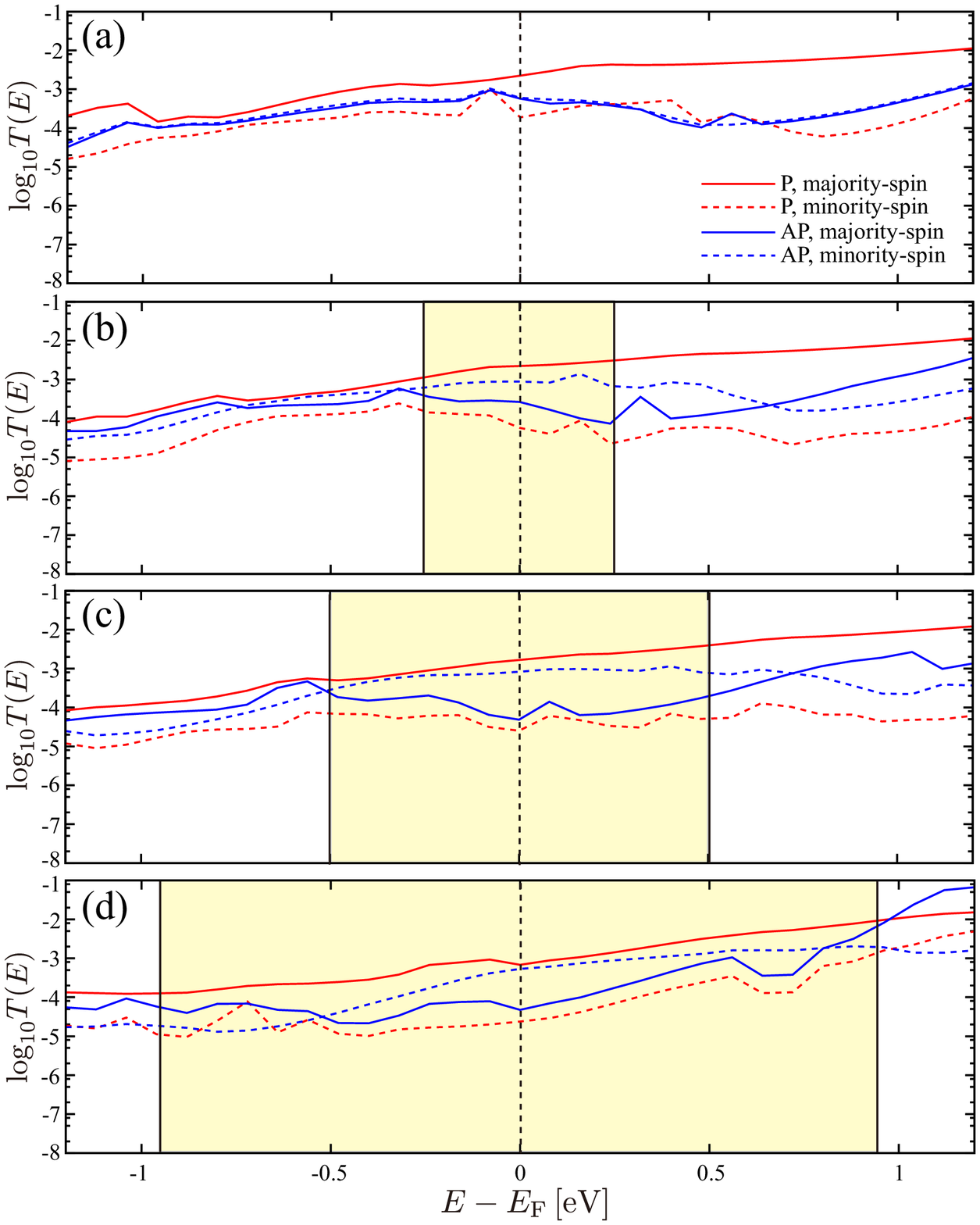}
\caption{\label{trans_MAO} The same as Fig. \ref{trans_MgO} for Fe/MgAl$_2$O$_4$/Fe(001) MTJs with (a) $V=0.0\,{\rm V}$, (b) $V=0.5\,{\rm V}$, (c) $V=1.0\,{\rm V}$, and (d) $V=1.9\,{\rm V}$.}
\end{figure}
Figures~\ref{I-V_V-MR_V-P}(a) and \ref{I-V_V-MR_V-P}(b) show the bias voltage dependencies of the currents in the MgO- and MgAl$_2$O$_4$-based MTJs, respectively. Here, $I_{\rm P}$ and $I_{\rm AP}$ represent the total (majority-spin plus minority-spin) currents in the case of the parallel and antiparallel magnetization of electrodes. Using these $I_{\rm P}$ and $I_{\rm AP}$, the optimistic MR ratio is defined by
\begin{eqnarray}
{\rm MR\, ratio}\,[\%]=100 \times \frac{I_{\rm P}-I_{\rm AP}}{I_{\rm AP}}\label{eq:MR}.
\end{eqnarray}
We show in Fig.~\ref{I-V_V-MR_V-P}(c) the bias voltage dependences of the MR ratios in the MgO- and MgAl$_2$O$_4$-based MTJs. We see that in both the MTJs, the MR ratio decreases as the bias voltage $V$ increases and eventually vanishes at the critical value $V_{\rm c}$. We can understand this behavior from Figs.~\ref{I-V_V-MR_V-P}(a) and \ref{I-V_V-MR_V-P}(b) and Eq. (\ref{eq:MR}) as follows. In a small bias region, the parallel current $I_{\rm P}$ has larger values than the antiparallel current $I_{\rm AP}$, leading to positive values of MR ratios. The value of the MR ratio decreases as $V$ increases, owing to the increase in $I_{\rm AP}$, as seen from Figs.~\ref{I-V_V-MR_V-P}(a) and \ref{I-V_V-MR_V-P}(b) and the denominator of Eq. (\ref{eq:MR}). As $V$ increases further, $I_{\rm AP}$ increases more rapidly and finally has the same value as $I_{\rm P}$ at $V_{\rm c}$, where the MR ratio becomes zero. Note that such a rapid increase of $I_{\rm AP}$ in $V>1.5\,{\rm V}$ is a common feature for both the MTJs, which is attributed to the enhancement of the electron tunneling between the $d_{3z^2-r^2}$-based majority- and minority-spin bands, whose lower band edges are located around $E-E_{\rm F}=-1.2\,{\rm eV}$ and $2\,{\rm eV}$ at $V=0\,{\rm V}$, respectively [see Figs. \ref{mgo_cond-band}(b) and \ref{mgo_cond-band}(d) for the MgO-based MTJ, and Figs. \ref{mao_cond-band}(b) and \ref{mao_cond-band}(d) for the MgAl$_2$O$_4$-based MTJ]. In Fig. \ref{I-V_V-MR_V-P}(d), we show the bias voltage dependences of the current polarization defined by $P\,[\%]=100 \times (I_{\rm P,maj}-I_{\rm P,min})/(I_{\rm P,maj}+I_{\rm P,min})$, where $I_{\rm P,maj}$ and $I_{\rm P,min}$ are the majority- and minority-spin currents in the case of parallel magnetization, respectively. In both the MTJs, the polarization sharply increases for small $V$ and has an almost constant value for intermediate $V$. For large $V$, the current smoothly decreases in both the MTJs.

Next, let us focus on the clear difference in $V_{\rm c}$ ($\sim$$0.3\, {\rm V}$) between the MgO- and MgAl$_2$O$_4$-based MTJs shown in Fig.~\ref{I-V_V-MR_V-P}(c). By comparing Figs.~\ref{I-V_V-MR_V-P}(a) and \ref{I-V_V-MR_V-P}(b), we find that this difference mainly comes from the different behavior of $I_{\rm P}$ in a high-$V$ region ($V>1\,{\rm V}$): the $I_{\rm P}$--$V$ curve maintains its slope beyond $V=1\,{\rm V}$ in the MgAl$_2$O$_4$ case, while it does not in the MgO case. To discuss such a difference in $I_{\rm P}$, we need to remember that the current is obtained by integrating the transmittance $T_{\sigma}(\epsilon)$ from $\epsilon=\mu_R$ to $\mu_L$, as shown in Eq. (\ref{eq:current}). In our calculations, since the origin of the energy $\epsilon$ was set to the average of the chemical potentials, $0=\mu_L+\mu_R$, the integration was carried out from $\epsilon=-{\rm e}\,V/2$ to ${\rm e}\,V/2$. Figures \ref{trans_MgO}(a)--\ref{trans_MgO}(d) show the energy dependences of the transmittance integrated over ${\bf k}_{\parallel}$ for various bias voltages in the MgO-based MTJ \cite{remark2}. Each shaded area represents each integral region. In Fig.~\ref{trans_MgO} (a), we can see a clear steplike structure at $E-E_{\rm F} \approx -1.2\,{\rm eV}$ in the majority-spin transmittance in the parallel magnetization case, whose integrated value provides the dominant contribution to $I_{\rm P}$. The steplike structure is identical to that at $E-E_{\rm F} \approx -1.2\,{\rm eV}$ in Fig~\ref{mgo_cond-band}(a), which reflects the band edge of the $d_{3z^2-r^2}$-based band leading to large values of transmittance, mentioned in the previous section. As $V$ increases, the steplike structure moves to higher energies, which is due to the upper shift of the density of states by ${\rm e}\,V/2$ in the left Fe electrode \cite{2009Rungger-PRB}. Around $1\,{\rm V}$, the structure is involved in the integral region as shown in Fig.~\ref{trans_MgO} (d). This causes the decrease in the slope of the $I_{\rm P}$--$V$ curve around $1\,{\rm V}$ [see Fig.~\ref{I-V_V-MR_V-P}(a)], which is the origin of the smaller $V_{\rm c}$. On the other hand, the majority-spin transmittance in the MgAl$_2$O$_4$-based MTJ with parallel magnetization of the Fe electrodes exhibits a very smooth energy dependence with no steplike structure as shown in Fig.~\ref{trans_MAO}(a), which corresponds to that in Fig.~\ref{mao_cond-band}(a). As explained in the previous section, this is because the $d_{3z^2-r^2}$-based band exists in the whole energy region around the Fermi level because of the band-folding effect in the Fe electrodes \cite{2012Miura-PRB,2012Sukegawa-PRB}. Owing to such a smooth energy dependence in the majority-spin transmittance, the slope of the $I_{\rm P}$--$V$ curve in Fig.~\ref{I-V_V-MR_V-P}(b) hardly changes, even when $V$ exceeds $1\,{\rm V}$, yielding the larger $V_{\rm c}$. From all these results and considerations, we can conclude that the difference in the critical bias $V_{\rm c}$ between MgO- and MgAl$_2$O$_4$-based MTJs mainly originates from the band-folding effect \cite{2012Miura-PRB,2012Sukegawa-PRB}, which occurs only in MgAl$_2$O$_4$-based MTJs.

Finally, we comment on the transmittance in the case of antiparallel magnetization of electrodes. Here, for simplicity, let us focus on the electron transmission from majority-spin bands in the left electrode to minority-spin bands in the right electrode, which is expressed by the blue solid curves in Figs.~\ref{trans_MgO} and \ref{trans_MAO}. In the MgO-based MTJ at zero bias, the majority-spin bands have the $d_{3z^2-r^2}$-based component in $E-E_{\rm F} \gtrsim -1.2\, {\rm eV}$, as shown in Fig.~\ref{mgo_cond-band}(b). On the other hand, the minority-spin bands have that component in $E-E_{\rm F} \gtrsim 2.0\, {\rm eV}$ and $E-E_{\rm F} \lesssim -2.4\, {\rm eV}$, as shown in Fig.~\ref{mgo_cond-band}(d). Therefore, electrons with the $d_{3z^2-r^2}$-based component can tunnel through the barrier in the shared region $E-E_{\rm F} \gtrsim 2.0\, {\rm eV}$, which is the main origin of the humplike structure of the transmittance at $E-E_{\rm F} \gtrsim 1.5\, {\rm eV}$ shown in Fig.~\ref{trans_MgO}(a). When the bias voltage is increased, the humplike structure moves to lower energies, as shown in Figs.~\ref{trans_MgO}(a)--\ref{trans_MgO}(d). This is because the shared energy region moves to lower energies due to the lower shift of the density of states in the right Fe electrode. In the case of MgAl$_2$O$_4$-based MTJ, relatively smooth energy dependencies with no specific features were found in the transmittances, as shown in Figs.~\ref{trans_MAO}(a)--\ref{trans_MAO}(d). This reflects the fact that the $d_{3z^2-r^2}$-based component exists in wide energy regions around the Fermi level in both the majority- and minority-spin bands, owing to the band-folding effect [see Figs.~\ref{mao_cond-band}(b) and \ref{mao_cond-band}(d)].

\section{summary}
We studied bias voltage effects on the spin-dependent transport properties of Fe/MgAl${}_2$O${}_4$/Fe(001) MTJs by means of the nonequilibrium Green's function method and the density functional theory. In particular, the bias voltage dependence of the MR ratio was investigated in detail by comparing it with that of the Fe/MgO/Fe(001) MTJ. In both MTJs, the MR ratio decreases as the bias voltage increases and eventually vanishes at a critical bias voltage $V_{\rm c}$. We also found that the MgAl${}_2$O${}_4$-based MTJ has a larger $V_{\rm c}$ than the MgO-based MTJ. In order to understand such a difference in $V_{\rm c}$, we analyzed the energy dependences of the current and transmittance for both the MTJs. From these results, it was revealed that the MgAl${}_2$O${}_4$-based MTJ with parallel magnetization of electrodes has a large majority-spin transmittance in a wide energy region around the Fermi level, owing to the band-folding effect in the electrodes, which is the origin of the larger $V_{\rm c}$. Such a robustness of the MR ratio against the bias voltage in Fe/MgAl${}_2$O${}_4$/Fe(001) MTJs is advantageous for device applications over MgO-based MTJs.
\begin{acknowledgments}
The authors are grateful to K. Hono, S. Mitani, S. Kasai, and H. Sukegawa for useful discussions and critical comments. This work was partly supported by TDK Corporation, by Grant-in-Aids for Scientific Research (S) (Grant No. 16H06332) and (B) (Grant No. 16H03852) from the Ministry of Education, Culture, Sports, Science and Technology, Japan, by NIMS MI2I, and also by the ImPACT Program of Council for Science, Technology and Innovation, Japan.
\end{acknowledgments}


\begin{thebibliography}{99}
\bibitem{note} Note that the lower limit of the color bar is eight orders larger than the MgO case.
\end{thebibliography}

\end{document}